# Intermolecular Correlations and Mean Square Relative Displacements in $C_{60}$ Fullerite

V.I. Zubov (a, b) and C.G. Rodrigues (c)

*(a) Instituto de Física, Universidade Federal de Goiás,C.P. 131, 74001 Goiânia-GO, Brazil*

*(b) Department of Theoretical Physics, People's Friendship University, Moscow, Russia*

*(c) Núcleo de Pesquisa em Física da Universidade Católica de Goiás, Goiânia -GO, Brazil*
cloves@pucgoias.edu.br

We study molecular properties of the high-temperature modification of fullerites. In the present work we calculate the intermolecular correlations and the mean square relative displacements in $C_{60}$. The Girifalco intermolecular potential is utilized. The calculations have been made in the whole interval from the equilibrium point with orientationally ordered phase up to the spinodal temperature. To take into account the lattice anharmonicity we use the correlative method of unsymmetrized self-consistent field. Its second order allows one to investigate correlations in fcc lattice between the nearest, second, third and fourth neighbors. The anharmonicity has strong effect on the intermolecular correlations at $T > 700$ K, causing their drastic rise near the spinodal point. The dependence of the correlation moments on the distance between molecules and on the crystallographic direction is considered. We compare our results with those for solid Ar and discuss the influence of a special feature of the interaction potentials on the intermolecular correlations and the mean square molecular displacements.

Мы изучаем молекулярные свойства высокотемпературных модификаций фуллеритов. В данной работе рассчитываются межмолекулярные корреляции и среднеквадратичные относительные смещения молекул в $C_{60}$. Используется межмолекулярный потенциал Жирифалко. Расчеты проводятся на всем интервале от температуры равновесия с ориентационно упорядоченной фазой до точки спинодали. Для учета ангармонизма применяется корреляционный метод несимметризованного самосогласованного пола. Его второй порядок позволяет изучать корреляции между ближайшими, вторыми, третьими и четвертыми соседями в ГЦК решетке. Ангармонизм сильно влияет на межмолекулярные корреляции при $T > 700$ K, вызывая их резкий рост вблизи точки спинодали. Рассматривается зависимость корреляционных моментов от расстояния между молекулами и кристаллографического направления. Результаты сравниваются с полученными для твердого аргона и обсуждается влияние особенностей потенциала взаимодействия на межмолекулярные корреляции и среднеквадратичные смещения.

## 1. Introduction

Since the discovery of the fullerenes [1], and especially after the elaboration of effective ways of their preparation [2] that gave rise to their production in sufficient quantities for the growth of crystals of macroscopic dimensions (the fullerites), these materials have been attracting a great attention from scientists [3–6]. The fullerenes of greatest abundance, $C_{60}$ and $C_{70}$ have been studied most intensively. Among other things, thermodynamic properties of fullerites have been investigated. Most of them are due to the lattice vibrations, whereas the dominant contribution to their specific heat comes from the intramolecular degrees of freedom. Using the correlative meth-od of unsymmetrized self-consistent field (CUSF) [7, 8], extensive theoretical studies have been performed on thermal and elastic properties of the high-temperature, orientationally disordered modifications of $C_{60}$ and $C_{70}$ fullerites, e.g. [9–12]. At $T > 700$ K, the lattice anharmonicity has a strong effect on their thermal expansion coefficient and elastic moduli.

In the present paper we calculate the quadratic correlation moments (QCM) between molecular displacements from the lattice points and their mean square relative displacements (MSRD) for $C_{60}$. Along with the lattice vibrational spectrum, these quantities are the most important dynamical characteristics of crystal lattices [13, 14]. Being closely related to fluctuations [15, 16], the QCM bear on important phenomena like the phase transitions, the thermodynamic stability, the spinodal decomposition, etc. The MSRD express the effective amplitudes of molecular vibrations which are measurable values. Lindemann's melting criterion [14] has been stated in terms of the mean square molecular displacements. Up to now, some interest has been shown in this criterion [17]. This is important for $C_{60}$ fullerite since discussions have been carried out recently on the possibility of its melting [18–21], see also [9, 10, 22].

## 2. Basic Equations

We consider a strongly anharmonic crystal with pairwise central interactions

$$U(\mathbf{r}_1, \mathbf{r}_2, ..., \mathbf{r}_N) = \tfrac{1}{2} \sum_{i \neq j} \Phi(|\mathbf{r}_i - \mathbf{r}_j|) \equiv \tfrac{1}{2} \sum_{i \neq j} \Phi(ij). \tag{1}$$

The mean square relative displacement of any two molecules $i$ and $j$ in a crystal can be written in the form

$$D_{ab}(ij) \equiv \overline{(q_{ia} - q_{jb})^2} = \overline{q_{ia}^2} + \overline{q_{jb}^2} - 2C_{ab}(ij). \tag{2}$$

Here $a$ and $b$ are the indices of the Cartesian components of the molecular displacements from the lattice points, and $C_{ab}(ij) = \overline{q_{ia} q_{jb}}$ the quadratic correlation moments. For perfect crystals, $\mathbf{q}_i = \mathbf{r}_i - \hat{A}\mathbf{n}_i$ where $\hat{A}$ is the lattice matrix and $\mathbf{n}$ are integral-component vectors. In this case, $\overline{q_{ia}^2} = \overline{q_{jb}^2} \equiv \overline{q_a^2}$. We shall use a coordinate system in which one of the axes, for instance the $X$-axis, runs through the centers of given molecules, then $C_{ab}(ij) = C_{aa}(\hat{A}(\mathbf{n}_i - \mathbf{n}_j))\, \delta_{ab}$.

To calculate statistical averages in the right-hand side of Eq. (2) for a strongly anharmonic crystal we represent its spatial probability density as

$$W(\mathbf{r}_1, \mathbf{r}_2, ..., \mathbf{r}_N) = C' \, e^{-U'/\Theta} \, W^0(\mathbf{r}_1, \mathbf{r}_2, ..., \mathbf{r}_N), \tag{3}$$

where, as usually, $\Theta$ is the absolute temperature multiplied by the Boltzmann constant. The zeroth approximation

$$W^0(\mathbf{r}_1, \mathbf{r}_2, ..., \mathbf{r}_N) = e^{-U^0/\Theta} / \int e^{-U^0/\Theta} \, d\mathbf{r}_1 ... d\mathbf{r}_N = \prod_i w_i(\mathbf{r}_i) = \prod_i w(\mathbf{r}_i - \hat{A}\mathbf{n}_i) \tag{4}$$

contains the sum of the self-consistent potentials of molecules:

$$U^0(\mathbf{r}_1, \mathbf{r}_2, ..., \mathbf{r}_N) = \sum_i u_i(\mathbf{r}_i) = \sum_i u(\mathbf{r}_i - \hat{A}\mathbf{n}_i); \tag{5}$$

$U'$ is the perturbing potential

$$U'(\mathbf{r}_1, \mathbf{r}_2, ..., \mathbf{r}_N) = U - U^0, \qquad (6)$$

and $C'$ the normalization constant.

As usual, the potential energies (1) and (5) are represented as power series of the molecular displacements. In both cases, we take into consideration their terms up to the fourth order since the higher ones are generally small [7, 8]. Because of this, the zeroth approximation (4), (5) includes the greater part of anharmonicity whereas (6) is a small correction and the probability density (3) together with its normalization constant $C'$ can be expanded in power series of $U'/\Theta$. Up to its second order [23, 24], we have

$$\begin{aligned}
\overline{q_{ia}^2} = \overline{a_i^2} + \frac{1}{2\Theta^2} \sum_k \Big\{ & \Phi_{\alpha\beta}(ik)\, \Phi_{\gamma\delta}(ik)\, \overline{\beta_k \delta_k}(\overline{a_i^2 \alpha_i \gamma_i} - \overline{a_i^2}\,\overline{\alpha_i \gamma_i}) \\
& + \frac{1}{4}\, \Phi_{\alpha\beta\gamma}(ik)\, \Phi_{\delta\varepsilon\xi}(ik)\, [\overline{\gamma_k \xi_k}(\overline{a_i \alpha_i \beta_i \delta_i \varepsilon_i} - \overline{a_i^2}\,\overline{\alpha_i \beta_i \delta_i \varepsilon_i}) \\
& - 2\overline{\delta_i \varepsilon_i}\,\overline{\gamma_i \xi_i}(\overline{a_i^2 \alpha_i \beta_i} - \overline{a_i^2}\,\overline{\alpha_i \beta_i}) + (\overline{a_i^2 \alpha_i \delta_i} - \overline{a_i^2}\,\overline{\alpha_i \delta_i})(\overline{\beta_k \gamma_k \varepsilon_k \xi_k} - \overline{\beta_k \gamma_k}\,\overline{\varepsilon_k \xi_k})] \\
& + \frac{1}{3}\, \Phi_{\alpha\beta}(ik)\, \Phi_{\gamma\delta\varepsilon\xi}(ik)\, [\overline{\beta_k \xi_k}(\overline{a_i^2 \alpha_i \gamma_i \delta_i \varepsilon_i} - \overline{a_i^2}\,\overline{\alpha_i \gamma_i \delta_i \varepsilon_i}) \\
& + \overline{\beta_k \delta_k \varepsilon_k \xi_k}(\overline{a_i^2 \alpha_i \gamma_i} - \overline{a_i^2}\,\overline{\alpha_i \gamma_i})] \Big\};
\end{aligned} \qquad (7)$$

$$\begin{aligned}
C_{ab}(ij) = & \frac{1}{\Theta}\, \Phi_{\alpha\beta}(ij)\, \overline{a_i \alpha_i}\,\overline{b_j \beta_j} + \frac{1}{6\Theta}\, \Phi_{\alpha\beta\gamma\delta}(ij)\, (\overline{a_i \alpha_i \gamma_i \delta_i}\,\overline{b_j \beta_j} + \overline{a_i \alpha_i}\,\overline{a_j \alpha_j \gamma_j \delta_j}) \\
& - \frac{1}{4\Theta^2}(\Phi_{\alpha\beta\gamma}(ij)\,\Phi_{\delta\varepsilon\xi} + \Phi_{\alpha\beta}\Phi_{\gamma\delta\varepsilon\xi}(ij))\,\overline{a_i \alpha_i \gamma_i \delta_i}\,\overline{b_j \beta_j \varepsilon_j \xi_j} \\
& + \frac{1}{4\Theta^2} \sum_k \{4\Phi_{\alpha\gamma}(ik)\, \Phi_{\beta\delta}(jk)\, \overline{a_i \alpha_i}\,\overline{b_j \beta_j}\,\overline{\gamma_k \delta_k} \\
& + \Phi_{\alpha\beta\gamma}(ik)\, \Phi_{\delta\varepsilon\xi}(jk)\, \overline{a_i \alpha_i}\,\overline{b_j \delta_j}(\overline{\beta_k \gamma_k \varepsilon_k \xi_k} - \overline{\beta_k \gamma_k}\,\overline{\varepsilon_k \xi_k}) \\
& + \Phi_{\alpha\beta\gamma}(ij)\, (\Phi_{\delta\varepsilon\xi}(jk)\, \overline{a_i \alpha_i}\,\overline{b_j \beta_j \gamma_j \delta_j} - \Phi_{\delta\varepsilon\xi}(ik)\, \overline{a_i \alpha_i \gamma_i \delta_i}\,\overline{b_j \beta_j})\,\overline{\varepsilon_k \xi_k}\} \\
& + \frac{1}{6\Theta^2} \sum_k \{\Phi_{\alpha\gamma}(ik)\, \Phi_{\beta\delta\varepsilon\xi}(jk)\, \overline{a_i \alpha_i}(\overline{b_j \beta_j}\,\overline{\gamma_k \delta_k \varepsilon_k \xi_k} + \overline{b_j \beta_j \delta_j \varepsilon_j}\,\overline{\gamma_k \xi_k}) \\
& + \Phi_{\beta\gamma}(jk)\, \Phi_{\alpha\delta\varepsilon\xi}(ik)\, \overline{b_j \beta_j}\,(\overline{a_i \alpha_i}\,\overline{\gamma_k \delta_k \varepsilon_k \xi_k} + \overline{a_i \alpha_i \delta_i \varepsilon_i}\,\overline{\gamma_k \xi_k})\}. \qquad (8)
\end{aligned}$$

Here

$$\Phi_{\alpha\beta...}(ik) = \left.\frac{\partial ...\, \Phi(|\mathbf{r}|)}{\partial x_\alpha\, \partial x_\beta \, ...}\right|_{\mathbf{r} = \hat{A}(\mathbf{n}_i - \mathbf{n}_j)}, \qquad (9)$$

and for brevity, we use the following designation for the moments of the zeroth-order distribution functions:

$$\overline{\alpha_i \beta_i ...} = \overline{q_{i\alpha} q_{i\beta} ...}^0 = \int q_{i\alpha} q_{i\beta} ...\, w_i(\mathbf{r}_i)\, d\mathbf{r}_i, \qquad (10)$$

the Greek indices being dummy. Generally, the summation extends over all $k$ except $k = i$, and for (8) except $k = j$ as well. If intermolecular forces are short-range, one can restrict it to the nearest, and sometimes to the second neighbors of the considered molecules.

In the case of perfect strongly anharmonic crystals with cubic lattices, the nonzero moments are determined by formulae

$$\overline{q_\alpha^2}^0 = \frac{1}{3}\overline{q^2}^0 = \frac{\beta\Theta}{3K_2}, \qquad (11)$$

$$\overline{q_\alpha^2 q_\beta^2}^0 = \frac{1}{15}(1+2\delta_{\alpha\beta})\overline{q^4}^0, \qquad \overline{q_\alpha^4 q_\beta^2}^0 = \frac{1}{35}(1+4\delta_{\alpha\beta})\overline{q^6}^0, \qquad (12)$$

$$\overline{q^{2l}}^0 = \frac{6(2n-1)\Theta}{K_4}\overline{q^{2l-4}}^0 - \left(\frac{6K_2}{K_4} - \frac{5}{3}\overline{q^2}^0\right)\overline{q^{2l-2}}^0, \qquad l=2,3, \qquad (13)$$

where

$$K_2 = \frac{1}{3}\sum_{\alpha=1}^{3}\sum_{\mathbf{n}\neq 0}\Phi_{\alpha^2}(|\hat{A}\mathbf{n}|); \qquad K_4 = \frac{1}{5}\sum_{\alpha,\beta=1}^{3}\sum_{\mathbf{n}\neq 0}\Phi_{\alpha^2\beta^2}(|\hat{A}\mathbf{n}|), \qquad (14)$$

are the second- and fourth-order force coefficients and $\beta(K_2\sqrt{3/\Theta K_4})$ is the solution of the transcendental equation [7]

$$\beta = 3X\frac{D_{-2.5}(X+5\beta/6X)}{D_{-1.5}(X+5\beta/6X)}, \qquad (15)$$

in which $D_\nu$ are the parabolic cylinder functions.

Here we apply the foregoing formulae to the high-temperature modification of $C_{60}$ fullerite that has the fcc lattice (Fig. 1). For the intermolecular forces we use the Girifalco potential [25]

$$\Phi_G(r) = -\alpha\left(\frac{1}{s(s-1)^3} + \frac{1}{s(s+1)^3} - \frac{2}{s^4}\right) + \beta\left(\frac{1}{s(s-1)^9} + \frac{1}{s(s+1)^9} - \frac{1}{s^{10}}\right), \qquad (16)$$

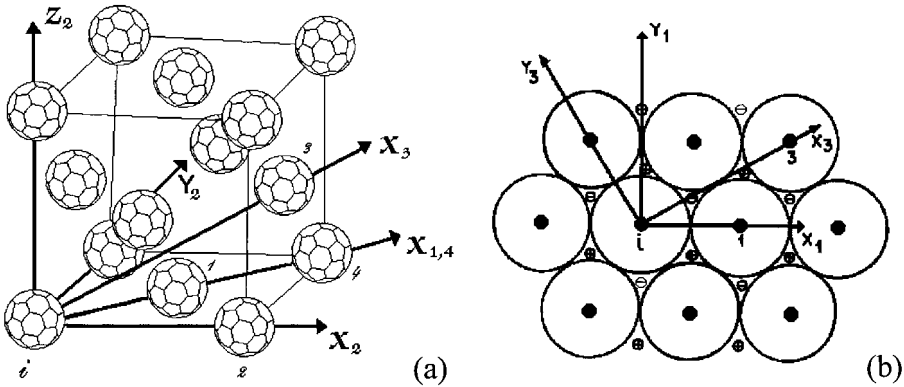

Fig. 1. The arrangement of the neighbours of a molecule in the high-temperature modification of $C_{60}$ fullerite with the coordinate axis for various molecular pairs: a) its spatial fragment, b) its section through a close-packed (111) plane

where $s = r/2a$, $a = 3.55 \times 10^{-8}$ cm, $\alpha = 7.494 \times 10^{-14}$ erg and $\beta = 1.3595 \times 10^{-16}$ erg. It has the minimum point $r_0 = 10.0558$ Å and the depth of the potential well is $\varepsilon/k = 3218.4$ K. For fcc crystals with short-range forces, such as (16), the expression (8) enables one to calculate the QCM up to the fourth neighbors. In Fig. 1a we show the $X$-axis for various molecular pairs and also other axes for the second neighbors. One can see that in all cases, the $X$-axis runs through the centers of the molecular pair, and hence, $C_{xx}$ is the longitudinal component of QCM. For the nearest and third neighbors we use coordinate systems whose $X$- and $Y$-axis lie in a close-packed (111) plane (Fig. 1b), and $Z$-axis is orthogonal to it. Evidently, for the second neighbors, the crystallographic coordinate system is the most convenient. The transversal components of the QCM between the fourth neighbors $C_{yy}(4) = C_{zz}(4)$ are independent of rotations around the $X$-axis.

## 3. Results of Calculations

We study QCM and MSRD for $C_{60}$ fullerite at normal pressure. We have calculated them along the lower, thermodynamically stable branch of the solution of the equation of state from the temperature of the orientational melting $T_{or} = 261.4$ K up to the spinodal point $T_s = 1917$ K [9]. To clear up the influence of the anharmonicity we also car-ried out such calculations in the harmonic and weakly anharmonic approximations. The case of weak anharmonicity follows from (7), (11)–(13) in the low-temperature limit when $X \gg 1$ and

$$\beta(X) \approx 3\left[1 - \frac{5}{X^2}\left(1 - \frac{21}{2X^2}\right)\right]. \tag{17}$$

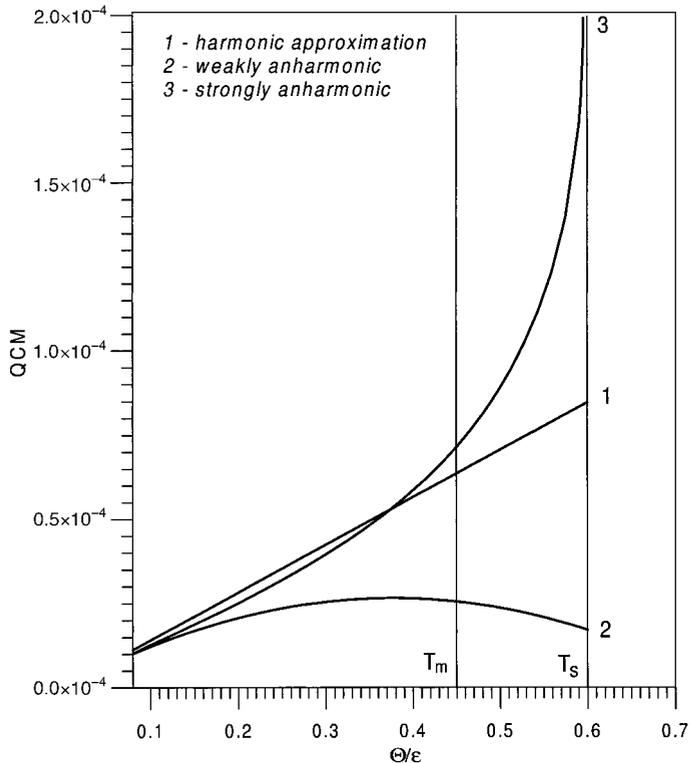

Fig. 2. QCM (dimensionless $C_{xx}(1)/a^2$) between longitudinal displacements of the nearest neighbors calculated using various approximations. $T_m$ and $T_s$ are the estimated [9, 10] melting temperature and spinodal point

The harmonic approximation takes place when $\beta \equiv 3$ and the third and fourth derivatives of the intermolecular potential are ignored.

In Fig. 2 we show results for the longitudinal correlation moment between the nearest neighbors $C_{xx}(1)$ obtained for the three cases. One can see that only at very low temperatures, all of them are very close one to another and both the anharmonic curves are convex upwards. At high temperatures the weakly anharmonic approximation for $C_{xx}(1)$ decreases with increasing temperature. As in the case of thermodynamic properties [9], the strong anharmonicity affects significantly the intermolecular correlations. It leads to the change of the convexity and to a sharp rise of this moment near the spinodal point $T \to T_s$. Such a rise of $C_{xx}(1)$ in the vicinity of $T_s$ agrees with the general idea for the behavior of fluctuations and correlations of physical quantities near the spinodal [15].

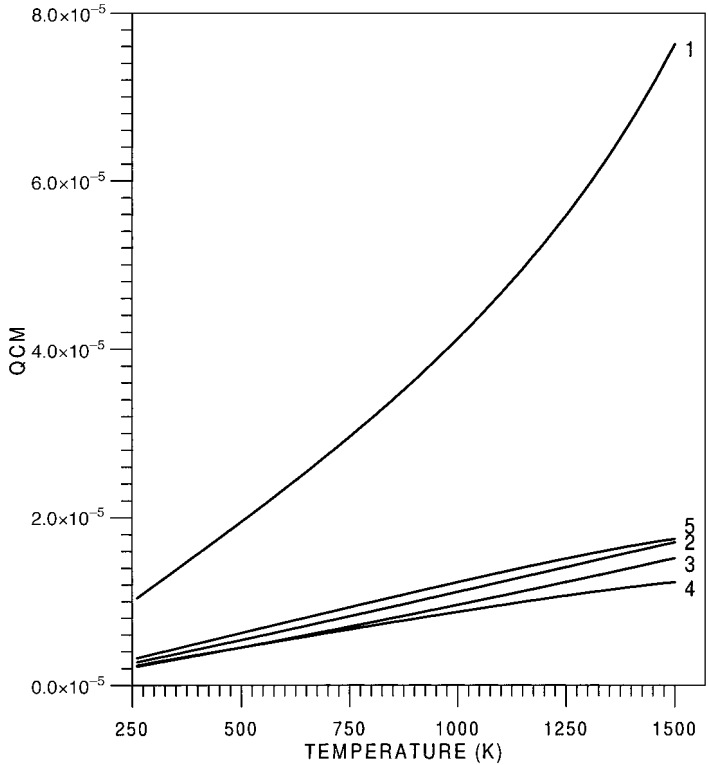

Fig. 3. All components of QCM between the nearest neighbors and longitudinal QCM between the third and fourth neighbors (in the dimensionless form $C_{aa}(n)/a^2$): (1) $C_{xx}(1)$, (2) $C_{yy}(1)$, (3) $C_{zz}(1)$, (4) $C_{xx}(3)$, (5) $C_{xx}(4)$

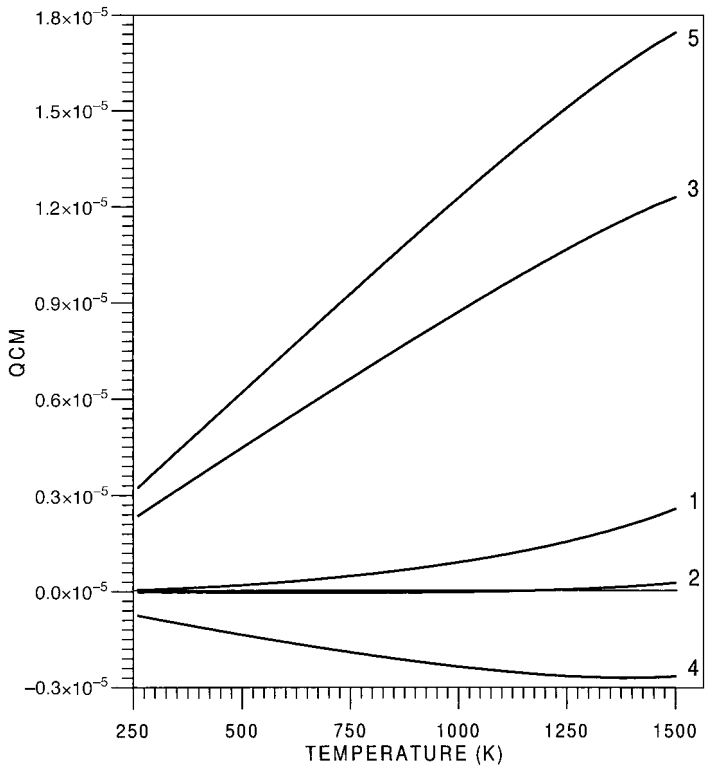

Fig. 4. Dimensionless QCM between non-nearest neighbors: (1) $C_{xx}(2)$, (2) $C_{yy}(2) = C_{zz}(2)$, (3) $C_{xx}(3)$, (4) $C_{yy}(3)$, (5) $C_{xx}(4)$

In Figs. 3–5 we present the results of our calculations of QCM and MSRD for the nearest, second, third and fourth neighbors. The distances between these pairs are $a$, $a\sqrt{2}$, $a\sqrt{3}$, and $2a$, respectively. We plot them up to the melting temperature $T_m$ estimated for normal pressure at 1450–1500 K [9, 10]. Note that it is close to the triple-point temperature evaluated by Caccamo [21] $T_{tr} \approx 1620$ K (see also [22]) that, in its turn, is in close agreement with our estimation [10] $T_{tr} \approx 1460$ K. Figure 3 demonstrates all components of the QCM between the nearest neighbors and the longitudinal QCM between the third and fourth neighbors. One can see that $C_{xx}(1)$ is at least three to four times greater than the other QCM. The remaining correlation moments are much less and can not be drawn in this scale.

In Fig. 4 we show all the QSM between non-nearest neighbors. The transversal moments between the second and fourth neighbors and also $C_{zz}(3)$ are not plotted since they are much less than the other ones. Note only that the inequalities

$$C_{xx}(1) > C_{xx}(4) > C_{yy}(1) > C_{zz}(1) > C_{xx}(3) > |C_{yy}(3)| > C_{xx}(2) > C_{yy}(2)$$
$$= C_{zz}(2) > |C_{zz}(3)| \approx |C_{yy}(4)| = |C_{zz}(4)|$$

are valid. The transversal QSM between the third and fourth neighbors are negative. This signifies that these pairs of molecules oscillate in these directions in almost opposite phases. For all considered pairs of molecules, the absolute values of the longitudinal

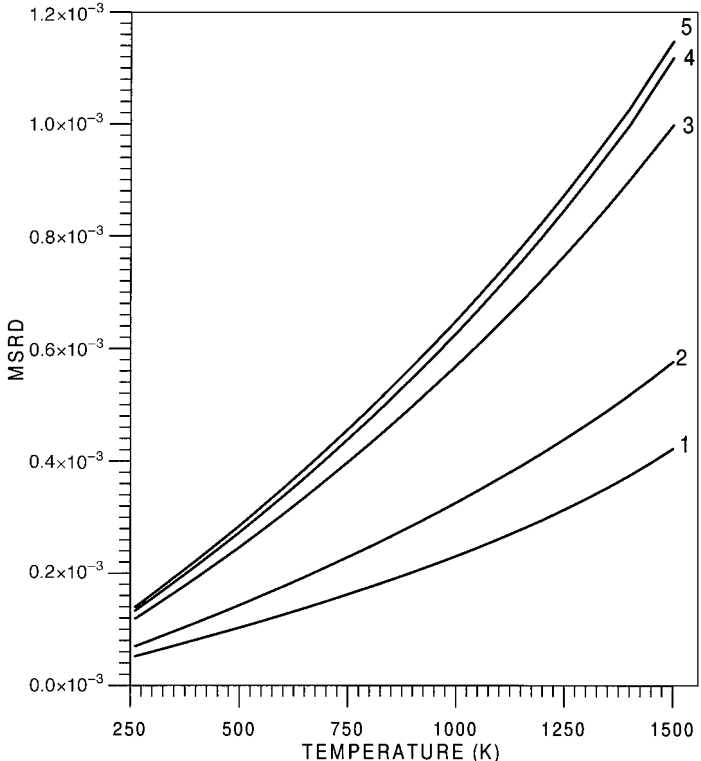

Fig. 5. Variance of the molecular position (1) in the zeroth approximation $\overline{q_a^2}^0/a^2$ (Eq. (11)) and (2) with corrections $\overline{q_a^2}/a^2$ (Eq. (7)), and MSRD: (3) $D_{xx}(1)/a^2$, (4) $D_{xx}(4)/a^2 \approx D_{xx}(3)/a^2$, and (5) $D_{xx}(2)/a^2 \approx D_{xx}(\infty)/a^2$

QCM are greater than those of the transversal ones. As a whole, they decrease rapidly with increasing distance between pairs. There is a strong anisotropy in this dependence.

In Fig. 5 we present the MSRD and, for comparison, the variance of the molecular position in the zeroth approximation (11) and with corrections (7). Only MSRD of the nearest, third and fourth neighbors differ noticeably from those of more distant ones. At $T_{tr} \approx 1460$ K [10], we obtain for the Lindemann parameter

$$\delta = \sqrt{\overline{\mathbf{q}^2}}\Big/a = \sqrt{3\overline{q_a^2}}\Big/a \qquad (18)$$

the value $\delta \approx 0.0405$. At the triple point temperature estimated in [21] (see also [22]), $T_{tr} \approx 1620$ K, we calculated $\delta \approx 0.0445$; and even at the estimation of 1800 K for this temperature [18], which in our opinion is overstated, $\delta \approx 0.0496$.

Note that for solid Ar which is the typical Van der Waals crystal with the same fcc lattice, the experimental data along its melting curve are [26] $0.084 \leq \delta \leq 0.097$, and the result of our calculation at normal pressure is 0.097. Hence the Lindemann parameter of $C_{60}$ fullerite is less than that of Ar by a factor 1.9–2.4.

A more detailed comparison between dynamical properties of these crystals is given below. In Fig. 6 are shown the longitudinal and transversal components of the QCM between the nearest neighbors. The MSRD between the nearest and distant molecular pairs appear in Fig. 7. One can see that in a qualitative sense, their behaviors are similar because these materials possess the same crystal lattice with short-range interactions. But their quantities for $C_{60}$ are far less than those for Ar.

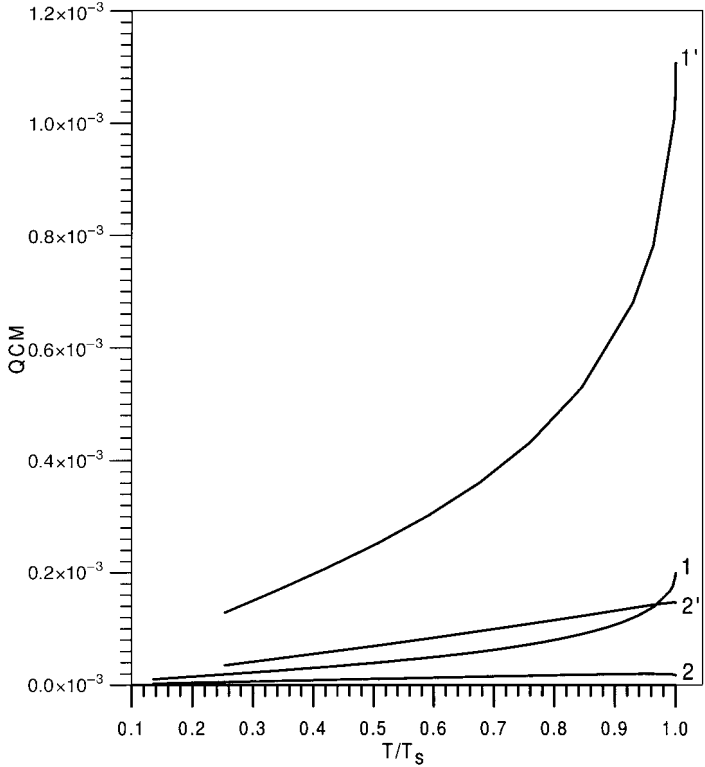

Fig. 6. Comparison between QCM in $C_{60}$ fullerite (1, 2) and solid Ar (1′, 2′): (1, 1′) $C_{xx}(1)/a^2$, (2, 2′) $C_{yy}(1)/a^2$

Like solid Ar, the high-temperature modification of $C_{60}$ fullerite is a van der Waals crystal [27], what is more, its anharmonicity is comparable with that of the rare gas solids. The only principal difference between their interaction forces is a peculiarity of the Girifalco potential for $C_{60}$, namely, the presence of a hard core of a finite radius. This allows one to conclude that such a distinction between results for $C_{60}$ and solid Ar is caused just by the presence of the hard core in the Girifalco potential. Note also that the relative intermolecular correlation $C_{xx}(1)/\bar{q}^2$ for $C_{60}$ is somewhat greater than that for Ar.

## 4. Conclusion

We have used the correlative method of unsymmetrized self-consistent field to study the intermolecular quadratic correlation moments and the mean square relative displacements in the high-temperature modification of $C_{60}$ fullerite taking into account the strong anharmonicity up to the fourth order. We compared our results with those obtained for solid Ar. From the above results, the following conclusions might be assumed:

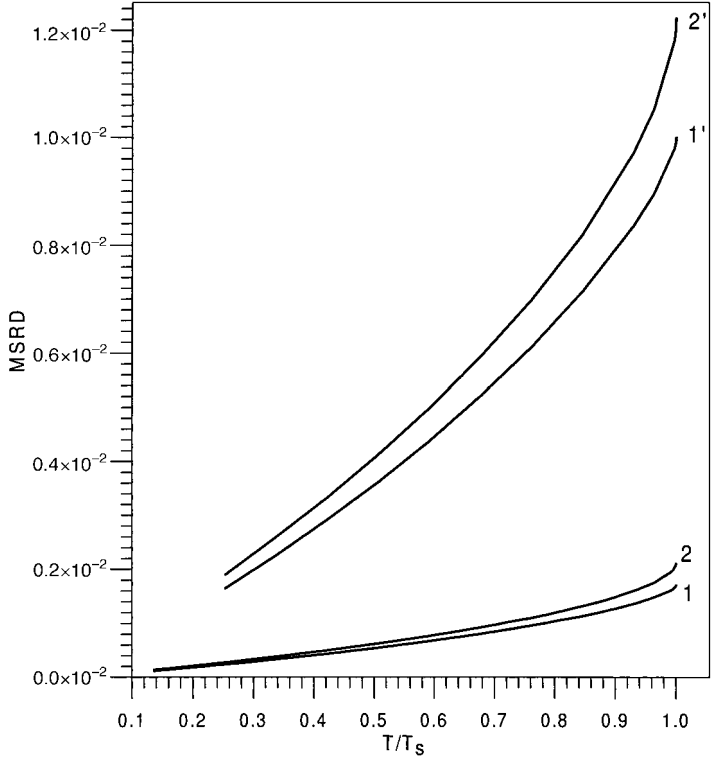

Fig. 7. Comparison between MSRD in $C_{60}$ (1, 2) and Ar (1′,2′): (1, 1′) $D_{xx}(1)/a^2$, (2, 2′) $D_{xx}(n)/a^2$

1. The strong anharmonicity of the lattice vibrations affects significantly the intermolecular QCM at high temperatures causing a drastic rise of $C_{xx}(1)$ near the spinodal point, which is in agreement with the general idea for the behavior of fluctuations and correlations of physical quantities near the spinodal.
2. The qualitative behavior of QCM and MSRD is governed by the type of the crystal lattice, whereas their magnitudes depend on the intermolecular forces. Because of this, the Lindemann parameter of $C_{60}$ fullerite is far less than that of simple van der Waals crystals, particularly, of Ar.
3. There is an appreciable anisotropy in the dependence of QCM on the intermolecular distance. In the fcc lattice they fall off more rapidly along the crystallographic axes and more slowly along the $\langle 110 \rangle$ directions.
4. Because of fairly small values of QCM, the MSRD between various molecular pairs in the fcc crystal are not too different from each other.

In conclusion, we notice that the obtained data may be used, for instance, in calculation of the possible melting curve of $C_{60}$ fullerite on the basis of Lindemann's melting criterion [14] because up to now attention has been paid to this criterion [17]. A comparison between results of such calculations and those of molecular dynamics simulations as well as between our results for the MSRD and experimental data in the future would be of interest.